\documentclass[twocolumn,amsmath,amssymb]{revtex4}

\usepackage{graphicx}
\usepackage{graphics}
\usepackage{dcolumn}
\usepackage{bm}

\def\BEq{\begin{equation}}
\def\EEq{\end{equation}}
\def\BEqA{\begin{eqnarray}}
\def\EEqA{\end{eqnarray}}
\def\BEn{\begin{enumerate}}
\def\EEn{\end{enumerate}}
\def\BWT{\begin{widetext}}
\def\EWT{\end{widetext}}

\def\a{\alpha}

\def\bra{\langle}

\def\ket{\rangle}

\begin{document}


\title{Quantum Zeno effect: A possible resolution to the leakage problem 
in superconducting quantum computing architectures}

\author{Andrei Galiautdinov}
 \affiliation{Department of Physics and Astronomy, 
University of Georgia, Athens, Georgia 30602}

\date{\today}

\begin{abstract}

We propose to use the continuous version of
the quantum Zeno effect to eliminate leakage to higher energy states 
in superconducting quantum computing architectures based on Josephson phase and flux 
qubits. We are particularly interested in the application of this approach to the 
single-step Greenberger-Horne-Zeilinger (GHZ) state protocol described in 
[A.\ Galiautdinov and J.\ M.\ Martinis, Phys. Rev. A {\bf 78}, 010305(R) (2008)].
While being conceptually appealing, the protocol was found to be plagued with a
number of spectral crowding and leakage problems. Here we argue that 
by coupling the qubits to a measuring device which continuously monitors 
leakage to higher energy states (say, to a very lossy resonator of 
frequency $\hbar\omega_{\rm dump} = E^{\rm qubit}_{3}-E^{\rm qubit}_{1}$, 
with 1 labeling the ground state of the qubit), we could potentially restrict the multi-qubit 
system's evolution to its computational subspace, thus circumventing the above 
mentioned problems.

\end{abstract}

\pacs{MORE...}

\maketitle


\section{Introduction}

The spectral crowding problem and the closely related problem of leakage to 
higher-energy states have pestered the field of superconducting quantum computing 
since its very inception \cite{Fazio1999}. For almost twenty years a tremendous amount 
of effort has been made to resolve both of these problems 
(see., e.\ g., 
\cite{JM2003, Montangero2007, Safaei2008, Rebentrost2009, Safaei2009, Gambeta2009,
Forney2010, Stojanovic2012, Schutjens2013, Motzoi2013, Martinis2016}).
The effort was certainly worth making, because every time an operation lifting 
the system off its ground state is performed (such as, e.\ g., a strong Rabi pulse), leakage 
out of the computational subspace develops, which results in the degradation of the 
gate fidelity. One may argue that even if the other problems facing quantum computing, 
such as scalability and decoherence, are ever successfully resolved, the crowding 
and the leakage problems would still be with us, simply as a matter of principle. 
As long as one insists on using Josephson junctions, one will have to deal with the matrix 
elements that kick the system all over the spectra of the corresponding 
anharmonic potentials.

As an example, consider the Greenberger-Horne-Zeilinger (GHZ) state 
protocol proposed in Ref.\ \cite{AGJM2008} for a
{\it fully connected network of three identical} Josephson phase qubits. 
In the rotating wave approximation (RWA) the system is described by the Hamiltonian, 
\begin{align}
H &= 
\vec{\Omega}_1 \cdot \vec{\sigma}_1 + \vec{\Omega}_2\cdot\vec{\sigma}_2
+\vec{\Omega}_3 \cdot \vec{\sigma}_3
\nonumber \\
& \quad 
+ \frac{1}{2}\bigg[g\left(\sigma_x^1 \sigma_x^2+\sigma_y^1 \sigma_y^2\right)
				+ \tilde{g}\sigma_z^1\sigma_z^2
\nonumber \\
& \quad \quad \quad \quad
+ g\left(\sigma_x^2 \sigma_x^3+\sigma_y^2 \sigma_y^3\right)
				+ \tilde{g}\sigma_z^2\sigma_z^3
\nonumber \\
& \quad \quad \quad \quad
+ g\left(\sigma_x^1 \sigma_x^3+\sigma_y^1 \sigma_y^3\right)
				+ \tilde{g}\sigma_z^1\sigma_z^3
 \bigg],
\end{align}
where $\Omega$'s are the Rabi frequencies, $g$ and $\tilde{g}$ are the coupling 
constants, and $\sigma$'s are the Pauli matrices.
The protocol consist of the following sequence of {\it symmetric} pulses:
\BEq
X_{\pi/2} U_{\rm int} Y_{\pi/2}|000\ket = e^{-i\a}e^{i(\pi/4)}|{\rm GHZ}\ket,
\EEq
with the entangling time set to $t_{\rm GHZ} = \pi/(2|g-\tilde{g}|)$,
and
\begin{align}
X_{\theta} &= X^{(3)}_{\theta}X^{(2)}_{\theta}X^{(1)}_{\theta},
\\
Y_{\theta} &= Y^{(3)}_{\theta}Y^{(2)}_{\theta}Y^{(1)}_{\theta},
\end{align}
being the simultaneous single-qubit rotations (the unimportant overall 
phase $\a$ depends on the values of $g$ and $\tilde{g}$). 
The crucial step in the protocol is the initial $Y$-rotation 
by $\pi/2$ performed on all the qubits in the circuit, which is supposed to 
result in the {\it fully uniform} superposition of the computational states,
\begin{align}
\label{eq:symmetricSTATE}
{|\psi\ket}_{\rm unif.} 
&=  (1/\sqrt{8})
\left( |000\ket + |001\ket+ \dots + |110\ket + |111\ket \right).
\end{align}
The protocol is very fast, but, unfortunately, is impossible to implement 
due to severe problems with spectral crowding \cite{MN2010}. The energy levels,
such as $|200\ket$, etc., quickly get populated during the initial rotations.

Since the problem of leakage out of the computational subspace is not going away
any time soon, one may try to tackle it by using the proverbial ``If you can't beat 
them, lead them'' principle. Since the higher energy levels are here 
to stay, we should look for ways to minimize their influence or make them irrelevant 
altogether. It seems that Nature herself, by way of quantum mechanical trickery, 
offers a curious possibility of doing just that, in the form of the so-called quantum 
Zeno effect \cite{Khalfin1957, MS1977, Itano1990, Panov1999, MITexperiment2006, 
Facchi2009, Matsuzaki2010, Kakuyanagi2015}. 
Before analyzing that effect in superconducting qubits, let us take a quick look at 
how it works in the simplest possible scenario.

\section{Brief review of the quantum Zeno effect}

Consider a two-level system, with basis states $|1\ket$ and $|2\ket$, whose 
dynamics is described by the Hamiltonian 
\begin{align}
H &= V (|1\ket\bra2| + |1\ket\bra2|) 
=
\begin{pmatrix}
0 & V \cr 
V & 0
\end{pmatrix},
\end{align}
where the off-diagonal matrix elements are chosen to be real, for convenience.
Denote the general time-dependent state of the system by
\BEq
|\psi(t)\ket = a_1(t)|1\ket + a_2(t)|2\ket, 
\EEq
subject to the normalization condition,
\BEq
\label{eq:3}
|a_1(t)|^2+|a_2(t)|^2=1.
\EEq
The evolution of the corresponding amplitudes, $a_1$ and $a_2$, 
is given by (here we use $\hbar =1$)
\begin{align}
i \frac{da_1}{dt}=Va_2, \quad
i \frac{da_2}{dt}=Va_1.
\end{align}
Assume,
\BEq
a_1(0)=1, \quad a_2(0) = 0.
\EEq
Then, in linear order, for small times, $t=\Delta t$, such that $|V \Delta t|\ll 1$, 
the amplitudes are given by
\BEq
a_1(\Delta t) \approx 1 + {o}(\Delta t^2), \quad
a_2(\Delta t) \approx -iV\Delta t,
\EEq
with the probability of finding the system in state $|2\ket$ being
\BEq
w_2(\Delta t) = |a_1(\Delta t)|^2 \approx V^2 \Delta t^2. 
\EEq
Normalization condition (\ref{eq:3}) then gives the probability of finding the 
system in state
$|1\ket$,
\BEq
w_1(\Delta t)=1-w_2(\Delta t)=1-V^2 \Delta t^2.
\EEq

Assume now that the total time, $T$, of system's evolution is divided into $n$ 
small equal time intervals, $\Delta t$, such that $T=n\Delta t$, with 
$|V\Delta t|\ll 1$. Assume additionally that at the end of the first time interval 
an ideal instantaneous measurement represented by the projection operator 
$P_2\equiv |2\ket\bra2|$ has been made -- typically implemented by some kind 
of tunneling process out of state $|2\ket$, -- that gave a {\it negative} result. 
Then the state of the system immediately after the measurement is described 
by the same ket $|1\ket$ in which the system was initially prepared at $t=0$.

We now calculate the probability $W^{(n)}_{1}(T)$ of the {\it complex event}, 
${\cal E}$, consisting of a series of $n$ $P_2$-measurements, resulting in the 
system {\it remaining} in the $|1\ket$ state at time $T$. For this to be possible, 
each of the intermediate measurements has to produce a {\it negative} result. 
Since all intermediate ``negative'' events are independent and each occurs with 
probability $w_1$, the overall probability of the complex event ${\cal E}$ is 
given by
\begin{align}
\label{eq:9}
W^{(n)}_{1}(T)&=[w_1(\Delta t)]^n 
\approx \left(1-V^2 \Delta t^2\right)^n
\nonumber \\
&= \left(1- \frac{V^2 T^2}{n^2}\right)^n 
\equiv \left(1- \frac{q}{n^2}\right)^n,
\end{align}
where $q \equiv V^2 T^2$.
In the limit $n\rightarrow \infty$, we get
\begin{align}
W^{(\infty)}_{1}(T)
&= \lim_{n\rightarrow \infty}
\left[\left(1- \frac{q}{n^2}\right)^{n^2/q}\right]^{q/n}
\nonumber \\
&=\lim_{n\rightarrow \infty}e^{-q/n}=1,
\end{align}
which constitutes the so-called quantum Zeno effect. Essentially, and 
counter-intuitively, by continually ``watching'' the system's $|2\ket$ state, 
we ``freeze'' the system in state $|1\ket$. 

\section{Application to a Rabi-driven superconducting qubit}

It is now clear how to use the quantum Zeno effect to restrict qubit's dynamics
to the computational subspace. By continually measuring the qubit's higher 
state(s), we will automatically confine the qubit to its computational subspace 
spanned by $|1\ket$ (ground state) and $|2\ket$, thus suppressing the leakage 
to states $|3\ket$, $|4\ket$, etc.

To be more specific, let us restrict consideration to a three-level system, 
\{$|1\ket$, $|2\ket$, $|3\ket$\}, with energies \{$E_1$, $E_2$, $E_3$\}, 
whose Rabi oscillation dynamics is implemented by a microwave drive of 
frequency, $\omega = \omega_{12}$, which is resonant with the 
$|1\ket \leftrightarrow |2\ket$ transition. This is described by the Hamiltonian, 
which, in the rotating wave approximation, is given by \cite{JM2003}
\begin{align}
H &= 
\begin{pmatrix}
0 & \Omega e^{i\phi} & 0 \cr 
\Omega e^{-i\phi} & 0 & \sqrt{2} \Omega e^{i\phi} \cr
0 & \sqrt{2} \Omega e^{-i\phi} & \eta \cr
\end{pmatrix},
\end{align}
where $\eta \equiv E_3-2E_2<0$ is the qubit anharmonicity 
(assuming $E_1=0$), $\phi$ is the phase of the drive, and $\Omega$ is 
the time-dependent (pulse-shaped) Rabi frequency, which we take to be 
constant, for simplicity. 

For future use, we will be interested in the case 
$\phi = -\pi/2$, which implements qubit rotations around the $Y$-axis of 
the Bloch sphere, with the corresponding Hamiltonian being
\begin{align}
\label{eq:12}
H &= 
\begin{pmatrix}
0 & -i\Omega & 0 \cr 
i\Omega & 0 & -i\sqrt{2} \Omega \cr
0 & i\sqrt{2} \Omega & \eta \cr
\end{pmatrix}.
\end{align}
Writing the unitary evolution operator in the form
\BEq
U(\Delta t) = 1 +(-i)H\Delta t + \frac{(-i)^2}{2!}H^2\Delta t^2 + \dots,
\EEq
and assuming that the initial state of the system is
\begin{align}
|\psi(0)\ket = a_{1}(0)|1\ket + a_{2}(0)|2\ket,
\\
|a_{1}(0)|^2+|a_{2}(0)|^2=1,
\\
a_3(0) = 0,
\end{align}
we find, in {\it second} order, the state at a later time $\Delta t$ to be
\begin{align}
\label{eq:15}
a_1(\Delta t)&= a_{1}(0)\left(1-\frac{1}{2}{\Omega}^2\Delta t^2\right) 
- a_{2}(0){\Omega} \Delta t,
\\
\label{eq:16}
a_2(\Delta t)&= a_{2}(0)\left(1-\frac{3}{2}{\Omega}^2\Delta t^2\right) 
+ a_{1}(0) {\Omega}\Delta t,
\\
\label{eq:17}
a_3(\Delta t)&=\sqrt{2}a_{2}(0) {\Omega}\Delta t +o(\Delta t^2). 
\end{align}
Correspondingly, the probabilities for finding the system in state $|3\ket$ and 
in the computational subspace are, respectively,
\begin{align}
w_3(\Delta t) &= |a_3(\Delta t)|^2 = 2|a_{2}(0)|^2 {\Omega}^2\Delta t^2,
\\
w_{\rm comp.}(\Delta t) &\equiv w_1(\Delta t)+w_2(\Delta t) 
= 1- 2|a_{2}(0)|^2 {\Omega}^2\Delta t^2.
\end{align}
By anology with the discussion around Eq.\ (\ref{eq:9}), in the case of a long 
sequence of $n$ $P_3$-measurements with negative outcomes, the probability 
$W^{(n)}_{{\rm comp.}}(T)$ of the complex event ${\cal E}_{\rm comp.}$, 
in which the qubit {\it remains in its computational subspace} at time $T$, can 
be estimated as,
\BWT
\begin{align}
\label{eq:22}
W^{(n)}_{{\rm comp.}}(T)
& \approx 
\left(1- 2|a_{2}|_{t=(n-1)\Delta T}^2 {\Omega}^2 \Delta t^2\right)
\dots
\left(1- 2|a_{2}|_{t=2\Delta T}^2 {\Omega}^2 \Delta t^2\right)
\left(1- 2|a_{2}|_{t=\Delta T}^2 {\Omega}^2 \Delta t^2\right)
\left(1- 2|a_{2}|_{t=0}^2 {\Omega}^2 \Delta t^2\right)
\\
& \geq 
\left(1- 2 {\Omega}^2 \Delta t^2\right)^n
= \left(1- \frac{2{\Omega}^2 T^2}{n^2}\right)^n \rightarrow 1, 
\quad
n\rightarrow \infty,
\end{align} 
\EWT
since $|a_2|^2\leq 1$ at every step of system's evolution.

As far as the evolution of the {\it amplitudes} is concerned, 
we return to Eqs.\ (\ref{eq:15}), (\ref{eq:16}), (\ref{eq:17}), and notice that 
if at time $\Delta t$, $|V\Delta t|\ll 1$, an ideal $P_3$-measurement is made 
with the negative outcome (for the system to be found in its $|3\ket$-state), 
the updated renormalized state of the system will be, in second order,
\begin{align}
a_1(\Delta t)&= a_{1}-a_{2}{\Omega}\Delta t 
- a_1\left(\frac{1}{2}-|a_2|^2\right){\Omega}^2\Delta t^2,
\\
a_2(\Delta t)&= a_{2}+a_{1}{\Omega}\Delta t 
- a_2\left(\frac{1}{2}+|a_1|^2\right){\Omega}^2\Delta t^2,
\\
a_3(\Delta t) & = 0,
\end{align}
where on the right hand side of each equation 
the label $(0)$ has been dropped for notational simplicity.
For $a_1=1$, $a_2=0$, we get
\begin{align}
a_1(\Delta t)&= 1-\frac{{\Omega}^2\Delta t^2}{2}\approx \cos (V\Delta t),
\\
a_2(\Delta t)&= {\Omega}\Delta t \approx \sin (V\Delta t),
\\
a_3(t) & = 0,
\end{align}
and for $a_1=0$, $a_2=1$, we get
\begin{align}
a_1(\Delta t)&= -{\Omega}\Delta t \approx -\sin ({\Omega}\Delta t),
\\
a_2(\Delta t)&= 1-\frac{{\Omega}^2\Delta t^2}{2}
\approx \cos ({\Omega}\Delta t),
\\
a_3(\Delta t) & = 0.
\end{align}
The above discussion indicates that in the presence 
of continuous higher-energy state mesurement the qubit will keep evolving 
as if the higher-energy state did not exist.

\begin{figure}[!ht]
\includegraphics[angle=0,width=1.00\linewidth]{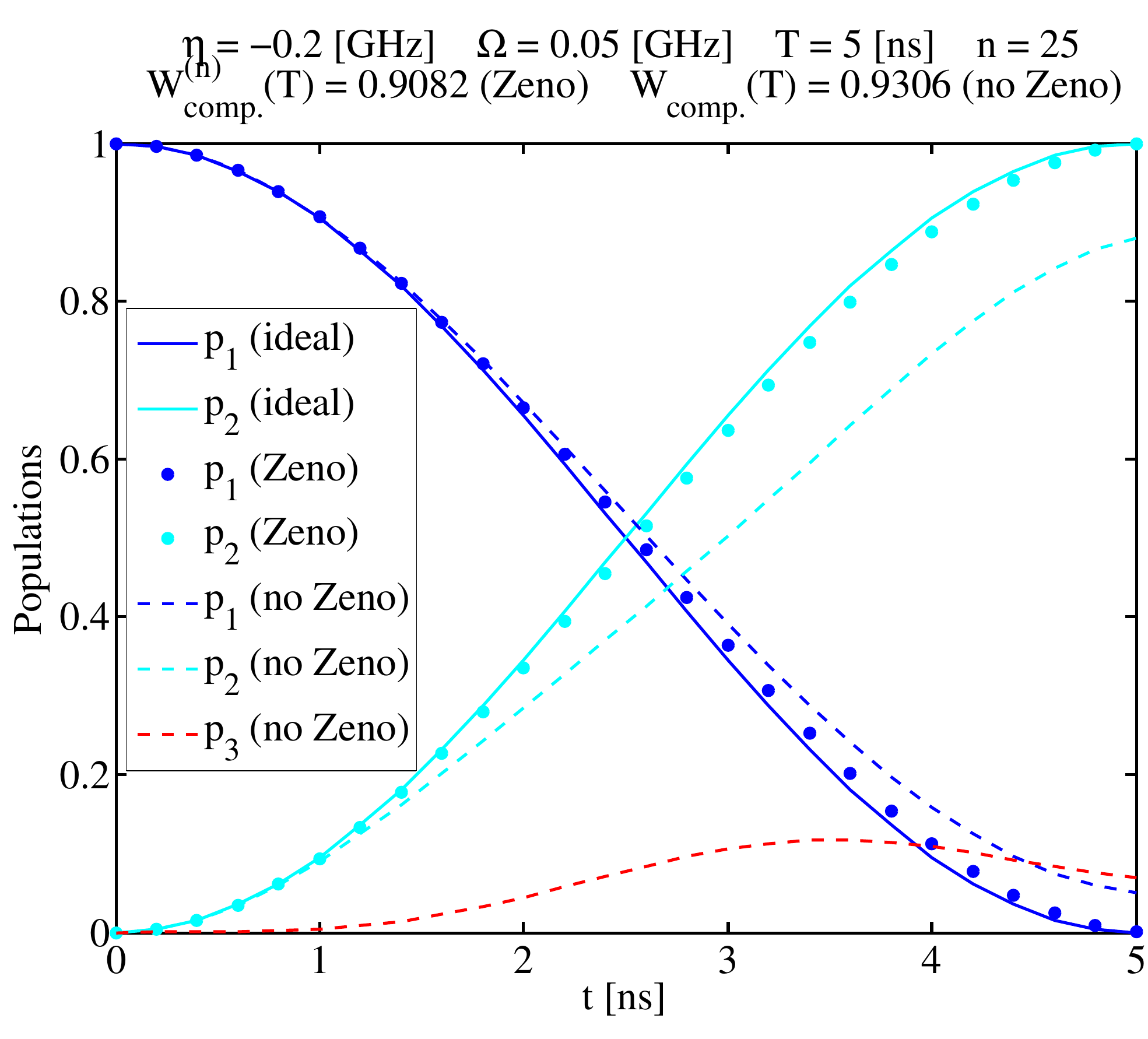}
\caption{ \label{fig:1} 
Populations of various qubit states during the Rabi pulse interrupted by 
$n=25$ projective measurements of state $|3\ket$.
}
\end{figure}

\begin{figure}[!ht]
\includegraphics[angle=0,width=1.00\linewidth]{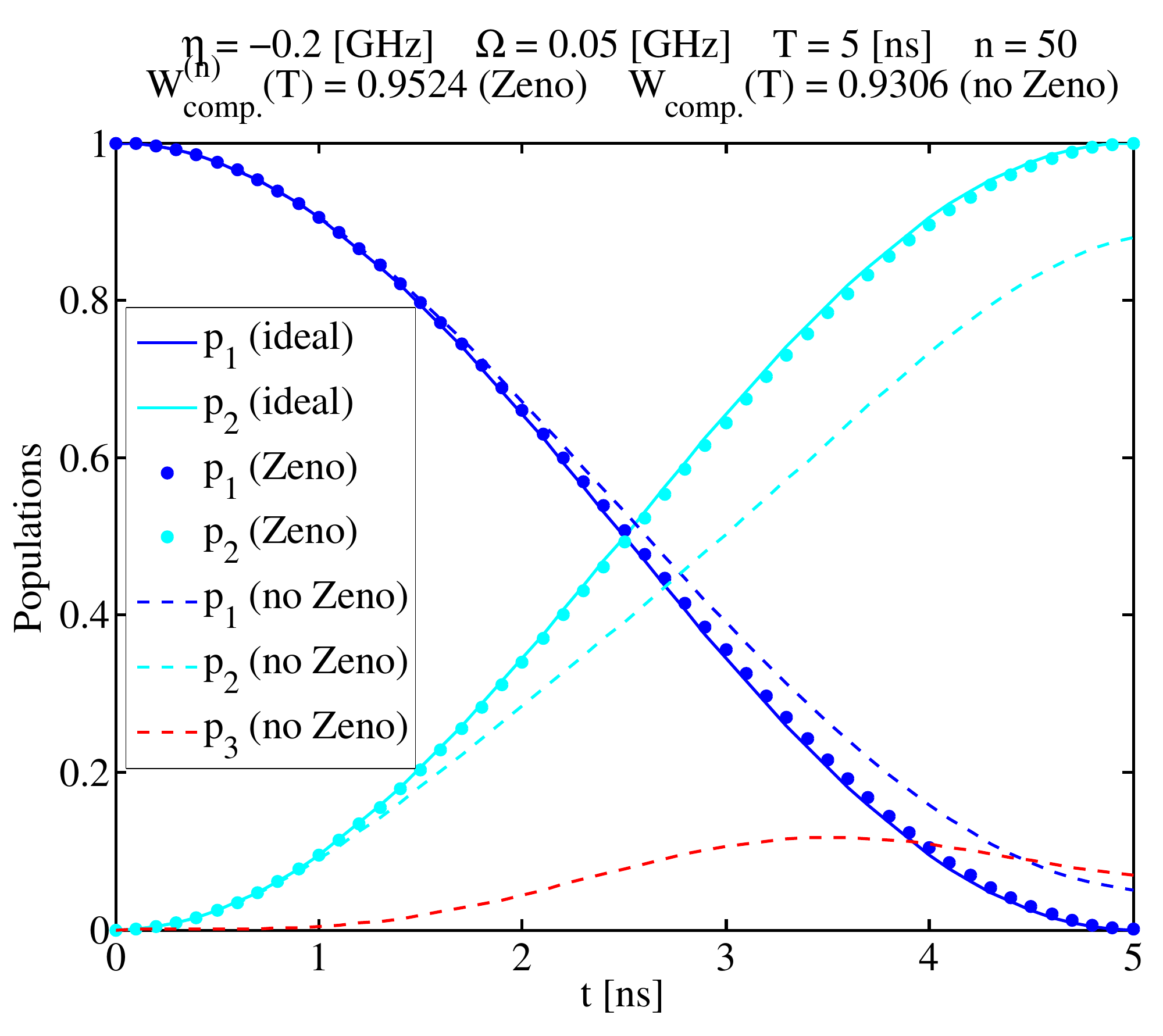}
\caption{ \label{fig:2} 
Populations of various qubit states during the Rabi pulse interrupted by 
$n=50$ projective measurements of state $|3\ket$.
}
\end{figure}

\begin{figure}[!ht]
\includegraphics[angle=0,width=1.00\linewidth]{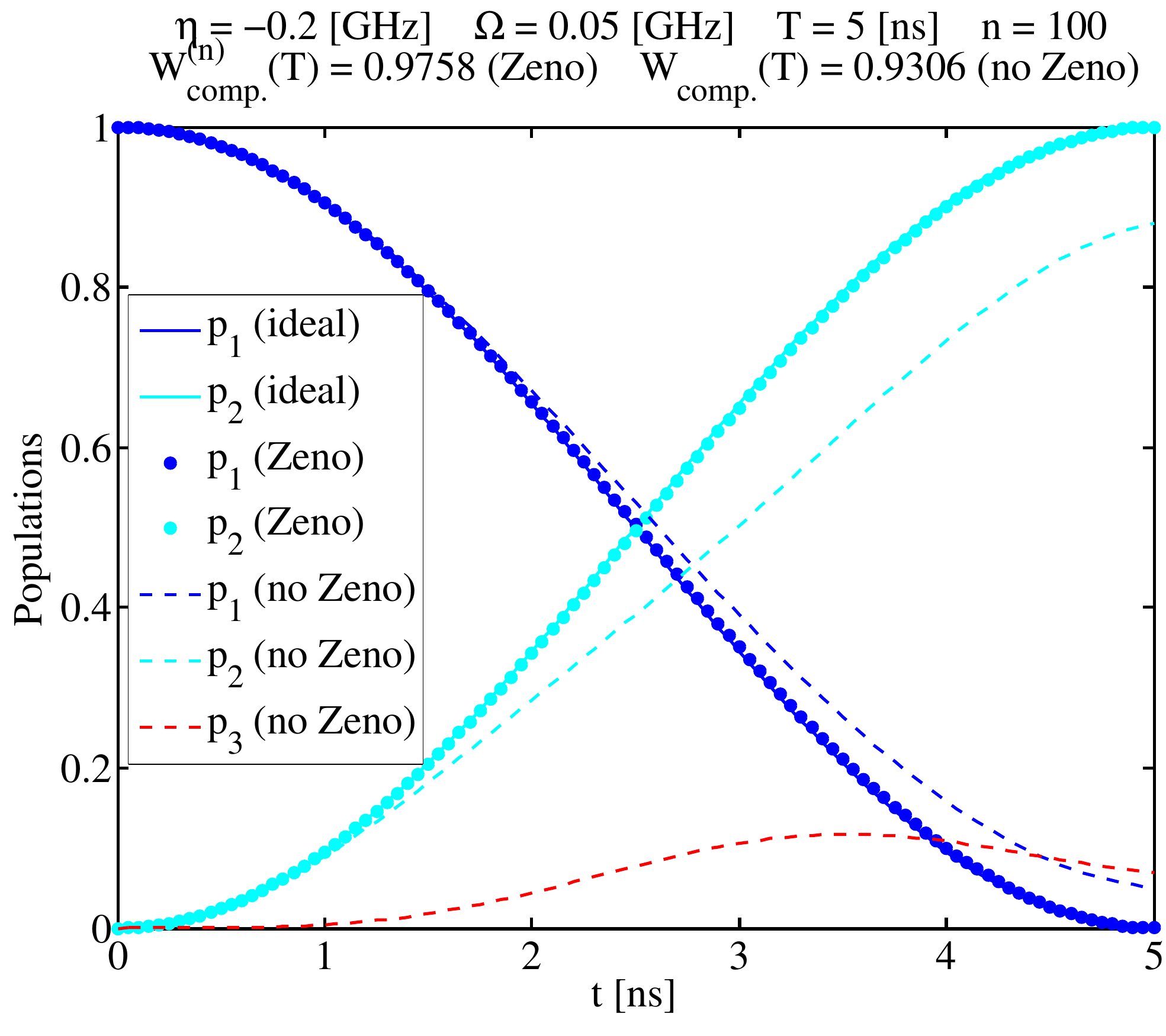}
\caption{ \label{fig:3} 
Populations of various qubit states during the Rabi pulse interrupted by 
$n=100$ projective measurements of state $|3\ket$.
}
\end{figure}

Figures \ref{fig:1}, \ref{fig:2}, and \ref{fig:3} show the time dependence of 
various state populations,
\BEq
p_i(t) \equiv |a_i(t)|^2, \quad i = 1,2,3,
\EEq
during the execution of the Rabi pulse applied to the system initially prepared
in its ground state, $|\psi(0)\ket = (1,0,0)$), and interrupted by $n=25$, 50, 
and 100 ideal $P_3$-measurements, with the assumption that the detector 
never clicked. We distinguish the following three cases:

The ideal case, shown just for comparison, corresponds to the evolution 
under the action of the model Hamiltonian,
\begin{align}
H_{\rm ideal} &= 
\begin{pmatrix}
0 & -i{\Omega} & 0 \cr 
i{\Omega} & 0 & 0 \cr
0 & 0 & \eta \cr
\end{pmatrix}.
\end{align}
without any leakage out of the computational subspace.

The Zeno case corresponds to the evolution under the action of $H$ given 
in (\ref{eq:12}) and subjected to $n$ projective $P_3$-measurements. The 
evolution was numerically calculated in accordance with the iterative scheme,
\begin{align}
|\psi_1\ket &= |1\ket ,
\nonumber \\
|\psi_k\ket &= 
\frac{P_{\rm comp.} |\psi_{k-1}\ket}
{\sqrt{||P_{\rm comp.} |\psi_{k-1}\ket||}} ,
\nonumber \\
|\psi_{k+1}\ket &= \exp\{-i H \Delta t\} |\psi_k\ket ,
\end{align}
for $k=2, 3, \dots, n$, where $P_{\rm comp.}$ is the projection operator 
onto the computational subspace,
\BEq
P_{\rm comp.} \equiv 1 - P_3 = 1-|3\ket\bra3|.
\EEq
At the final time $T=n\Delta t$ the qubit remains in its computational 
subspace (that is, $p_{3}(T)\equiv |a_3(T)|^2=0$), with the probability 
(let us call it the {\it survival probability}) given by the product 
(compare to (\ref{eq:22})),
\begin{align}
W^{(n)}_{{\rm comp.}}(T)
& = \prod_{k=1}^{n} \left(1- p_3(t_k)\right).
\end{align} 

Finally, the no-Zeno case corresponds to the exact evolution,
\BEq
U(t) = \exp\{-iHt\},
\EEq
with the leakage to $|3\ket$ taken into account. The probability for the 
system to be found in the computational subspace in a single 
measurement performed at the end of the evolution is 
\BEq
W_{\rm comp.}(T)= 1 - p_3(T).
\EEq 

A quick glance at the figures shows the presence of the Zeno effect in our 
system. There is a critical number of the measuring steps, $n_{\rm crit}$, 
depending on the choice of the system parameters, 
above which the survival probability $W^{(n)}_{{\rm comp.}}(T)$ exceeds 
$W_{\rm comp.}(T)$. In the limit $n\rightarrow \infty$, 
$W^{(n)}_{{\rm comp.}}(T)$ approaches 1, as expected.


\section{Taking a closer look at Eq.\ (\ref{eq:17}) and designing a 
measurement model}

\begin{figure}
\includegraphics[angle=0,width=1.00\linewidth]{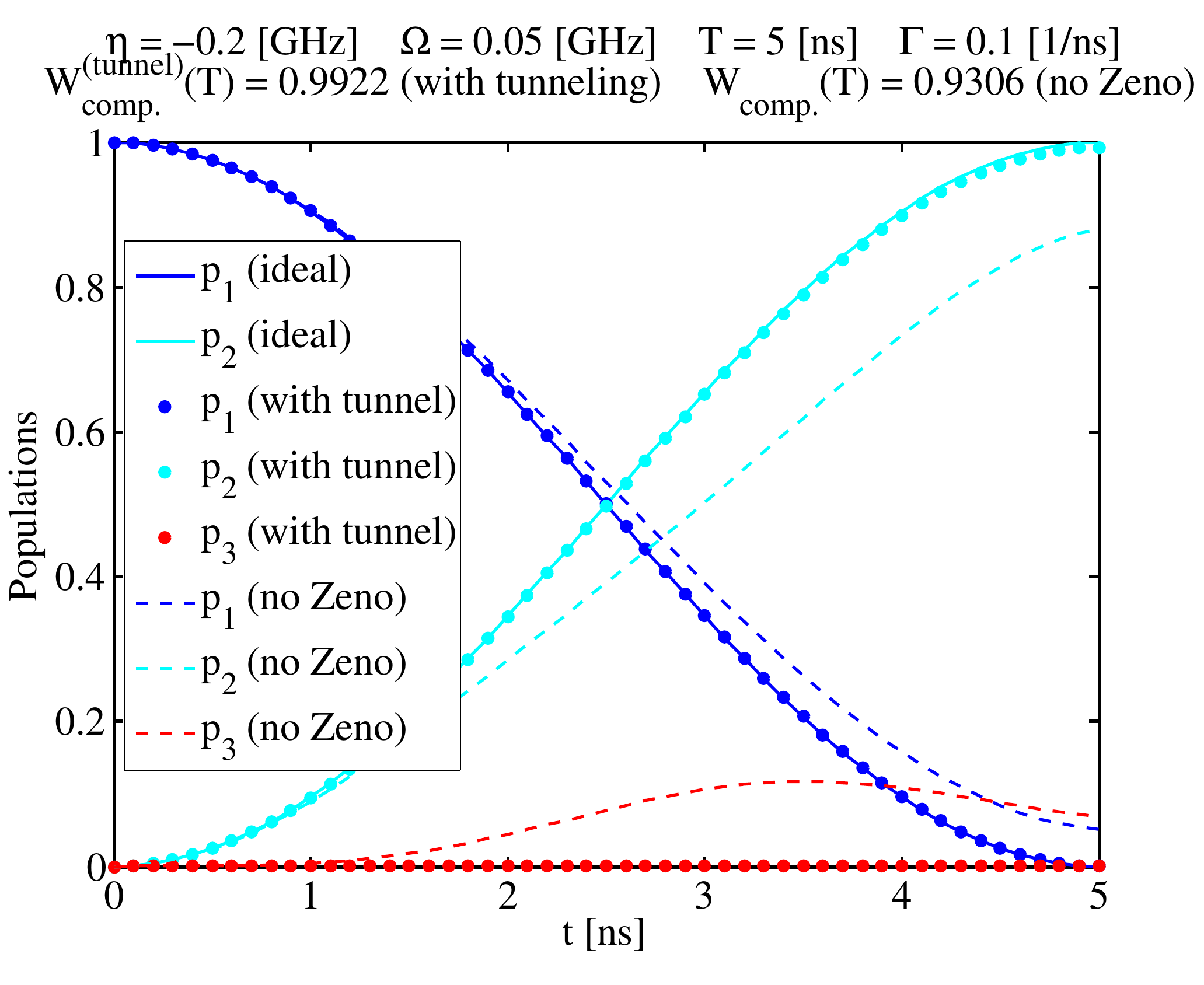}
\caption{ \label{fig:4} 
Populations of various qubit states during the Rabi pulse with the state $|3\ket$
subjected to tunneling.
}
\end{figure}

We now take a closer look at Eq.\ (\ref{eq:17}), in which the second order 
term was ignored. Restoring that $o(\Delta t^2)$ term, we find
\begin{align}
a_3(\Delta t) &= \sqrt{2}a_{2}(0) {\Omega}\Delta t 
+ \frac{\sqrt{2}}{2}\left[a_1(0) \Omega^2 - a_2(0)\Omega\eta i\right] 
\Delta t^2.
\end{align}
Assume now that the $|3\ket$ level is allowed to tunnel out with the rate 
$\Gamma$. The third diagonal matrix element of the Hamiltonian (\ref{eq:12}) 
would then get modified as
\BEq
\eta \rightarrow \eta - i\Gamma/2,
\EEq
giving
\begin{align}
\label{eq:42}
a_3(\Delta t) &= \sqrt{2}a_{2}(0) {\Omega}\Delta t 
\nonumber \\
&+ \frac{\sqrt{2}}{2}\left[a_1(0) \Omega^2 
- a_2(0) \Omega \left(\frac{\Gamma}{2}+\eta i\right)\right]  \Delta t^2.
\end{align}
We now make the following observation: if the tunneling rate is chosen to be
\BEq
\Gamma = \frac{4}{\Delta t},
\EEq
the first order term, $\sqrt{2}a_{2}(0) {\Omega}\Delta t$, in Eq.\ (\ref{eq:42}) 
would get canceled, giving
\begin{align}
\label{eq:49}
a_3(\Delta t) &=
\frac{\sqrt{2}}{2}\left[a_1(0) \Omega^2 - a_2(0)\Omega \eta i\right]
\Delta t^2,
\end{align}
which would result in the {\it fourth} order population of the unwanted state 
$|3\ket$.

Figure \ref{fig:4} shows the result of exact simulation for the system initially 
prepared in the ground state, using the Hamiltonian
\begin{align}
\label{eq:50}
H &= 
\begin{pmatrix}
0 & -i\Omega & 0 \cr 
i\Omega & 0 & -i\sqrt{2} \Omega \cr
0 & i\sqrt{2} \Omega & \eta - i\Gamma/2 \cr
\end{pmatrix},
\end{align}
with $\Gamma = 40$ [ns$^{-1}$] and $\Omega = 0.05$ [GHz] 
(this effectively corresponds to $n=50$ with $\Delta t = 0.1$ [ns] of the Zeno 
case). The survival probability is defined by
\BEq
W^{({\rm tunnel.})}_{{\rm comp.}}(t) = |a_1(t)|^2+|a_2(t)|^2.
\EEq
The corresponding dynamics can be viewed as a reasonable measurement model 
for state $|3\ket$. We see that for the chosen system's parameters leakage out of the 
computational subspace is virtually nonexistent.

\newpage
\section{Conclusions}

Working with the RWA Hamiltonian of a three-level qubit driven by a Rabi 
pulse, we showed that, by allowing the system to continuously tunnel out 
of its higher-energy state (the greater, $\Gamma$ the better), we effectively 
confine system's dynamics to the computational subspace. Such quantum 
Zeno behavior can potentially be used to resolve the long standing spectral 
crowding problem that plagued the field of superconducting quantum 
computing for almost two decades.

%
%
%

\end{document}